\documentclass[12pt]{article}
\usepackage{sbc-template}
\usepackage{graphicx,url}
\usepackage{amsmath}
\usepackage{color}
\usepackage{wrapfig}
\usepackage{quoting}
\usepackage{subfigure}
\usepackage[utf8]{inputenc}

\title{User-Centric Health Data Using Self-sovereign Identities\footnote{This research is part of the INCT of the Future Internet for Smart Cities funded by CNPq, proc. 465446/2014-0, CAPES proc. 88887.136422/2017-00, and FAPESP, proc. 2014/50937-1.}}

\author{Alexandre Siqueira\inst{1}, Arlindo Flavio da Conceição\inst{1}, Vladimir Rocha\inst{2}}
\address{
  Universidade Federal de São Paulo (UNIFESP)
  \nextinstitute
  Universidade Federal do ABC (UFABC)
  \email{\{alexandre.siqueira, arlindo.conceicao\}@unifesp.br, }
  \email{vladimir.rocha@ufabc.edu.br}
}

\begin{document} 

\maketitle

\begin{abstract}
This article presents the potential use of the Self-Sovereign Identities (SSI), combining with Distributed Ledger Technologies (DLT), to improve the privacy and control of health data. The paper presents the SSI technology, lists the prominent use cases of decentralized identities in the health area, and discusses an effective blockchain-based architecture. The main contributions of the article are: (i) mapping SSI general and abstract concepts, e.g., issuers and holders, to the health domain concepts, e.g., physicians and patients; (ii) creating a correspondence between the SSI interactions, e.g., issue and verify a credential, and the US standardized set of health use cases; (iii) presenting and instantiating an architecture to deal with the use cases mentioned, effectively organizing the data in a user-centric way, that uses well-known SSI and Blockchain technologies.

\end{abstract}

\section{Introduction}
\label{intro}
Personal health records reveal a conflict between privacy protection laws and the ability to share personal information. Throughout life, people interact with doctors and healthcare service providers countless times, either for routine appointments or medical treatment. In these interactions, clinical information is obtained that could constitute a significant batch of medical records for future use. However, a single patient might have his/her medical data spread across several healthcare service providers, and thus create siloed databases that, in the current state --- fragmented, are useless for clinical application outside those silos. Added to this scenario, the increasing adoption of wearable devices, for collecting information about health and lifestyle, is creating a new patient-generated health data silo.

The fragmented nature of personal medical records prevents someone from having a unique and usable person's health track record that could help physicians make an accurate diagnosis. Thus, new methods are needed for creating a decentralized data structure, controlled by the users, to maintain their medical records. This appears to be the pathway to enabling health providers to store medical information while granting data ownership to the rightful owners --- the patients.

In health domain, the problem of data ownership can be overcome by the use of Decentralized Digital Identities, Decentralized Ledger Technologies, and Self-sovereign Identity. Decentralized Digital Identities (DIDs)~\cite{did2021} are a set of attributes that uniquely identify and represents an entity (e.g., patients) in a digital context. The Self-sovereign Identity (SSI) are designed to enable entities 
to control its DIDs, using cryptographic tools, such as digital signatures~\cite{lacchain2020}. With that, users in the health domain could use these identifiers to interact among them (exchanging their identities), in a secure and reliable way, without loosing the data ownership. Decentralized Ledger Technologies (DLTs), such as Blockchain, had been investigated by Kassab \textit{et al.}~\cite{kassab2019} and Houtan \textit{et al.}~\cite{Houtan_2020} as core mechanisms for electronic health record systems. The use of this kind of technology and identities could be a means of confronting the challenge of sharing and securing sensitive medical information among healthcare parties, as well as ensuring patients maintain sovereignty over their data.



In the remaining of this article, Section~\ref{related} discusses related work. Section~\ref{ssi} presents the principles of SSI. Section~\ref{usecases} lists some use cases of SSI in medical data and personal health data. Section~\ref{arch} proposes the architecture based on SSI and DLT frameworks. Section~\ref{discuss} discusses the practical open problems related to the application of SSI in healthcare. Finally, in Section~\ref{conclusions} we present ideas for future works and our conclusions.

\section{Related work}
\label{related}

In recent years, several authors explored the idea of using intelligent agents in healthcare context to provide interoperability~\cite{Isern2016, CORTES2015, Wimmer2014}. More recently, some authors explored blockchain technology~\cite{shrier2016office, liu2016medical, Conceicao_2018}, most of them proposing strategies to improve mobility and security in EHRs using distributed ledgers.

Regarding using both SSI and DLT in the healthcare context, to the best of our knowledge, we found few works directly related to ours, two of them discussed below.

Bouras \textit{et al}.~\cite{bouras2020} discuss several aspects to achieve a decentralized identity management, using blockchain, in health systems. Among those aspects, they identify the key players, their roles, and the different health scenarios in which SSI and DLT technologies can be applied. In the first aspect, stakeholders (key players) were divided into: regulators (e.g., government), representatives (e.g., insurance and equipment companies), providers (e.g., physicians and nurses), and consumers (e.g., patients). In the second aspect, all players have the same roles (Issuer, Inspector, Holder), but with different functionalities, depending on the context. In these roles: the Issuer is responsible for creating claims; the Inspector is responsible for verifying the claims; the Holder is responsible for storing and control the claims. In the third aspect, the scenarios are divided into data exchanging, online pharmacy, payment, and research projects. In these scenarios, the authors identify the players associated and the requirements necessary (such as interoperability, data standards, anonymity, scalability, etc.) to deploy them in the real world. Besides these aspects, the work also lists several decentralized identity management solutions that use the SSI and DLT technologies, but only one of them (Evernym) has the Healthcare domain as an industry target.


Houtan \textit{et al}.~\cite{Houtan_2020} as well as the aforementioned work, also discuss on how the SSI and Blockchain can be applied in the health domain. The difference is in the focus on the challenges that arise when creating a healthcare information system (HIS) that must be decentralized, private, and secure in order to exchange patients' data. Among the challenges in using  DLTs, the authors identify that the blockchain category (e.g., public, private, hybrid), smart contract programming language, and consensus protocol create trade-off that must be analyzed by the architects and designers in order to avoid functional and non-functional misbehavior or security problems. Among the challenges to maintain the system private and secure, in the SSI context, they identify that the data collection (source of the data), interoperability, and access scenarios could lead to reveal some information to unauthorized parties (information leakage) or to lose information because the data is controlled and owned by the users. This work presented and analyzed several solutions using blockchain as the core for creating HIS, but concludes that most of them are not ready to be deployed in real-world scenarios.


\section{Self-Sovereign Identity (SSI)}
\label{ssi}

A person walks into a bar and asks for a drink. The bartender then asks to show some identification to confirm that he/she is old enough to buy liquor. The person presents his/her state-issued driver's license, containing the name, photo, date of birth, mother's name, and other personal information.  The bartender recognizes the presented document, notices the resemblance between the photo and the person, figures the persons' age by his date of birth, and, finally, confirms that the person can place the order. 

The described scenario, proposed by Windley~\cite{windley2018}, depicts examples of identification (the person presents his/her document to the bartender), authentication (the document proofs that the person is who he claims to be), and authorization (the person is old enough to buy a drink). Ordinary interactions like this are the inspiration behind a concept called self-sovereign identity. Windley describes:

\begin{quoting}[rightmargin=0cm,leftmargin=3cm]
\noindent
\textit{``Self-sovereign identity starts with the notion that we all are the makers of our own identity, online and off. Because they do not rely on any centralized authority, self-sovereign identity systems are decentralized, mirroring the way identity works in real life.''}~\cite{windley2018}
\end{quoting}

The SSI represent the user-centric identity model because it grants the users control over their data. User-centric approaches for managing digital identities date back to 2005 when Josang described a model where users would store their identifiers and credentials in a portable hardware device secured by a local secret, like a PIN~\cite{josang2005}. 
The advancements in portable devices and distributed ledger technologies brought improvements to the user-centric model, enabling the creation of two essential elements for SSI~\cite{lacchain2020}:

\begin{itemize}
    \item \textbf{Decentralized ledgers:} distributed records structures that store cryptographic proofs such as digital signatures and timestamps, allowing anyone to verify digital credentials issued by entities without the need of a central source of trust.
    \item \textbf{Digital wallets:} portable and secure personal repositories that allow users to manage identities and verifiable credentials within the phones, completely protected and under their control~\cite{lacchain2020}. It also enables voluntary information disclosure situations: users can select what information to disclose to whom. 
\end{itemize}

\section{Use cases of SSI in health applications}
\label{usecases}

There are several use cases in which SSI could be applied in the health domain~\cite{bouras2020}. The selected use cases described below were adapted from the United States Core Data for Interoperability (USCDI) standards~\cite{uscdi2020}.

\subsection{Allergies and intolerances}

\textbf{Context:} allergies and intolerances represent harmful or undesirable physiological responses presented by patients when exposed to a particular substance. Allergens can group into the following categories: drugs (e.g. penicillin or aspirin), food (e.g. milk or gluten), environmental (e.g. dust or mold), and non-medication (e.g. latex). 

\begin{wrapfigure}{L}{0.3\textwidth}
  \begin{center}
    \includegraphics[width=0.30\textwidth]{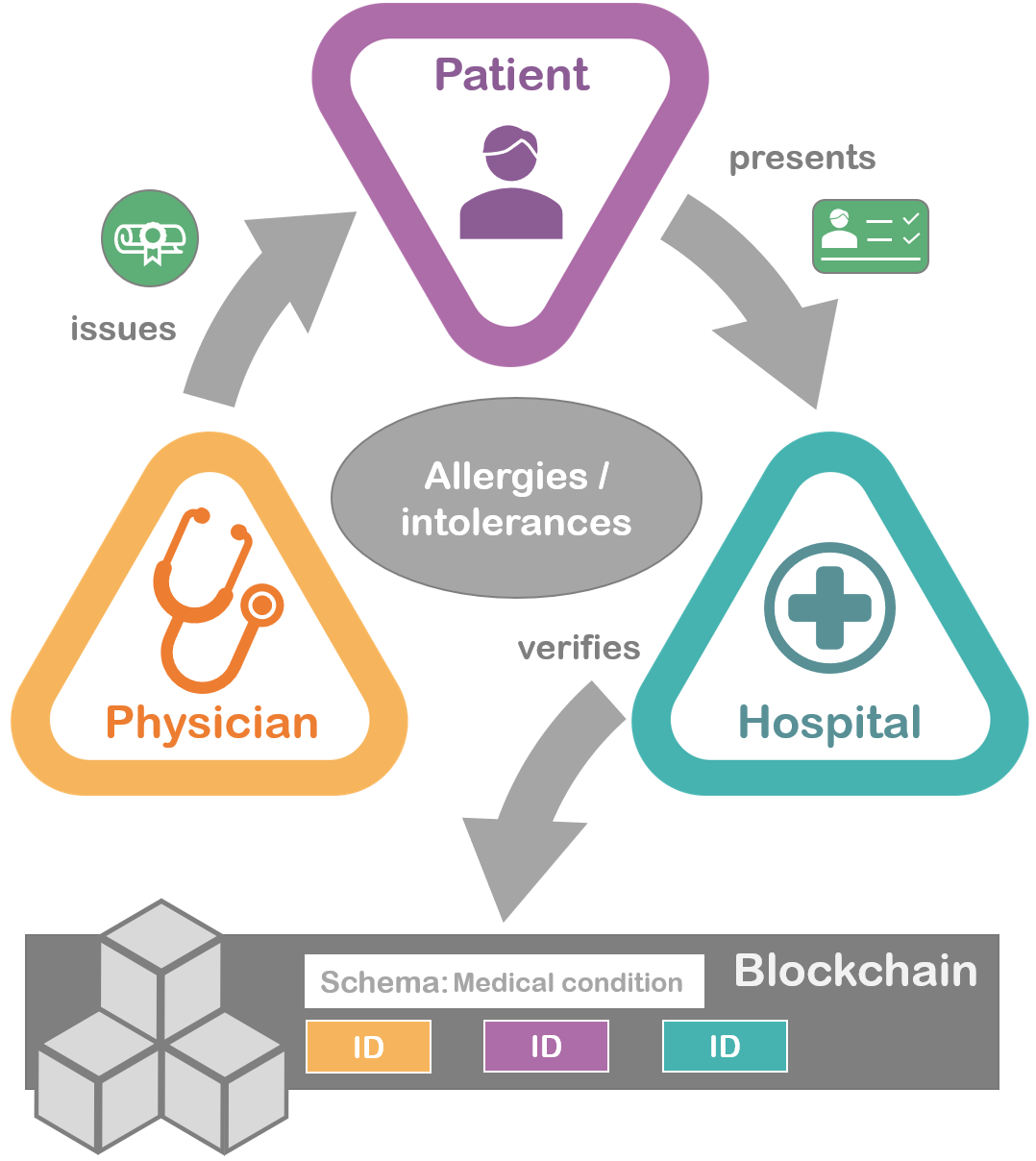}
  \end{center}
  \caption{Allergies and intolerances}
\end{wrapfigure}

\textbf{Problem/Challenge:} checking for existing allergies and intolerances is essential to provide proper care for patients.  Failing to capture this information during admission in the healthcare system can cause severe harm or even death. Allergic and intolerant patients usually inform of their known conditions, but an unconscious patient  cannot warn his/her physician.

\textbf{Storytelling}: Lisa wakes up with a severe menstrual cramps. She goes to the hospital and asks for something for the pain. The physician asks Lisa to present her medical conditions history, looking specifically for existing 
allergies, and current medications. Lisa uses her digital wallet to create a verifiable presentation with the information requested. While examining the claims, the physician notices she is allergic to a substance common in painkillers. Therefore, he searches for another medication to alleviate her pain. 

\textbf{Benefits}: using verifiable credentials to present known conditions for healthcare providers helps to automate specific inquiries that impact the diagnosis and treatment of patients without having to ask them explicitly.

\subsection{Clinical notes}

\begin{wrapfigure}{L}{0.3\textwidth}
  \begin{center}
    \includegraphics[width=0.30\textwidth]{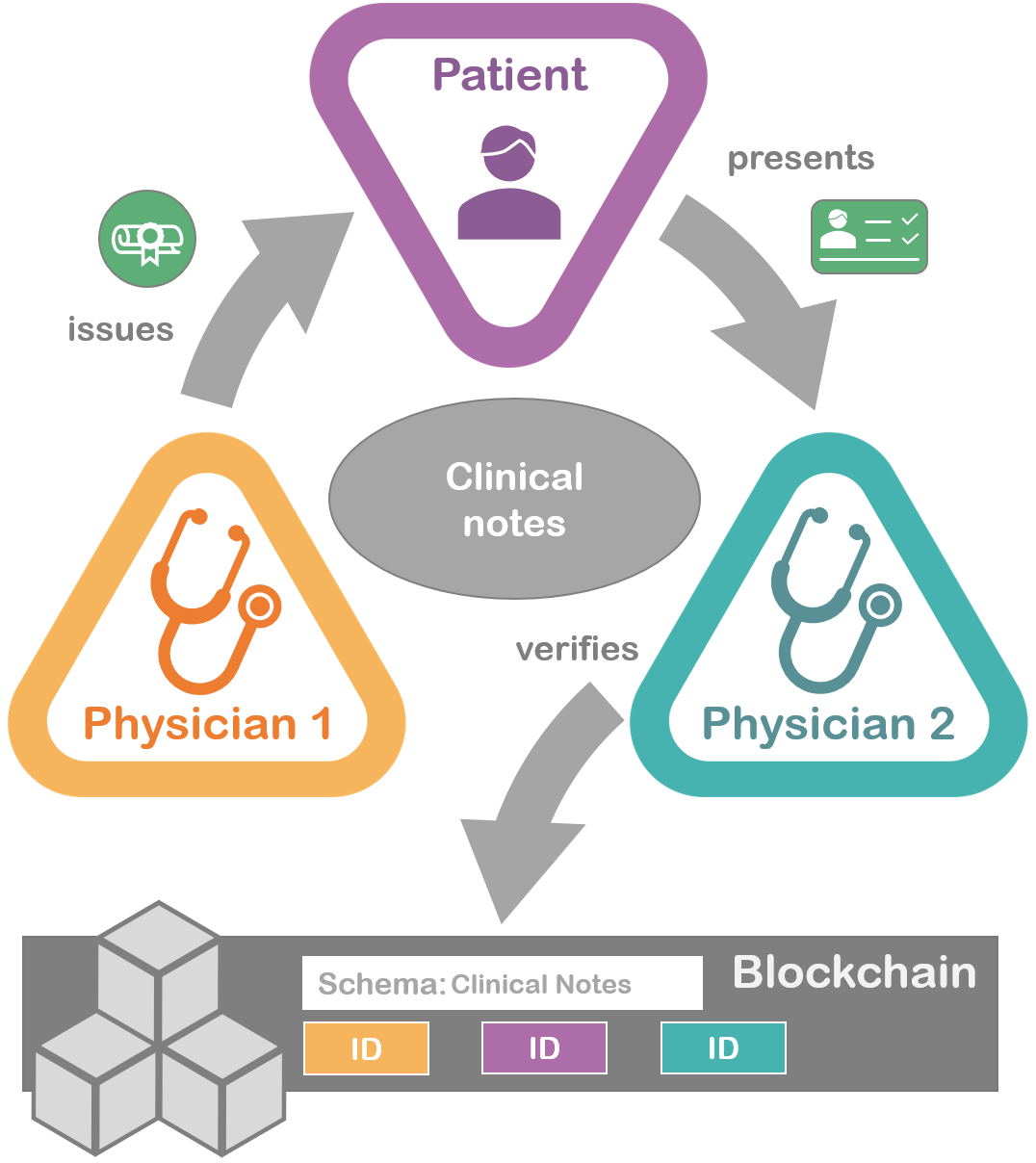}
  \end{center}
  \caption{Clinical Notes}
\end{wrapfigure}

\textbf{Context:} clinical notes are records taken during a medical evaluation~\cite{uscdi2020}. Those records can be structured (checklist) or unstructured (free text) data and may comprise items as consultation notes, history \& physical notes, etc.  

\textbf{Problem/Challenge:} clinical notes are the most frequent, most sensitive, and the least normalized pieces of patients' health records. Keeping this data into a personal record that can be shared privately with a healthcare provider would improve the quality of the health assessment while respecting patient privacy.

\textbf{Storytelling:} Charles goes on an appointment with a new cardiologist. He is seeking a second opinion about a health assessment he received from another physician. The cardiologist reviews the past medical evaluation from Charles' electronic health record and discover that the previous physician did not consider further lab tests to support his assessment. 

\textbf{Benefits:} addressing clinical notes as claims in an SSI digital wallet allows patients to consolidate their health history, 
breaking the silos of information. 

\subsection{Immunizations}

\begin{wrapfigure}{L}{0.3\textwidth}
 \begin{center}
    \includegraphics[width=0.30\textwidth]{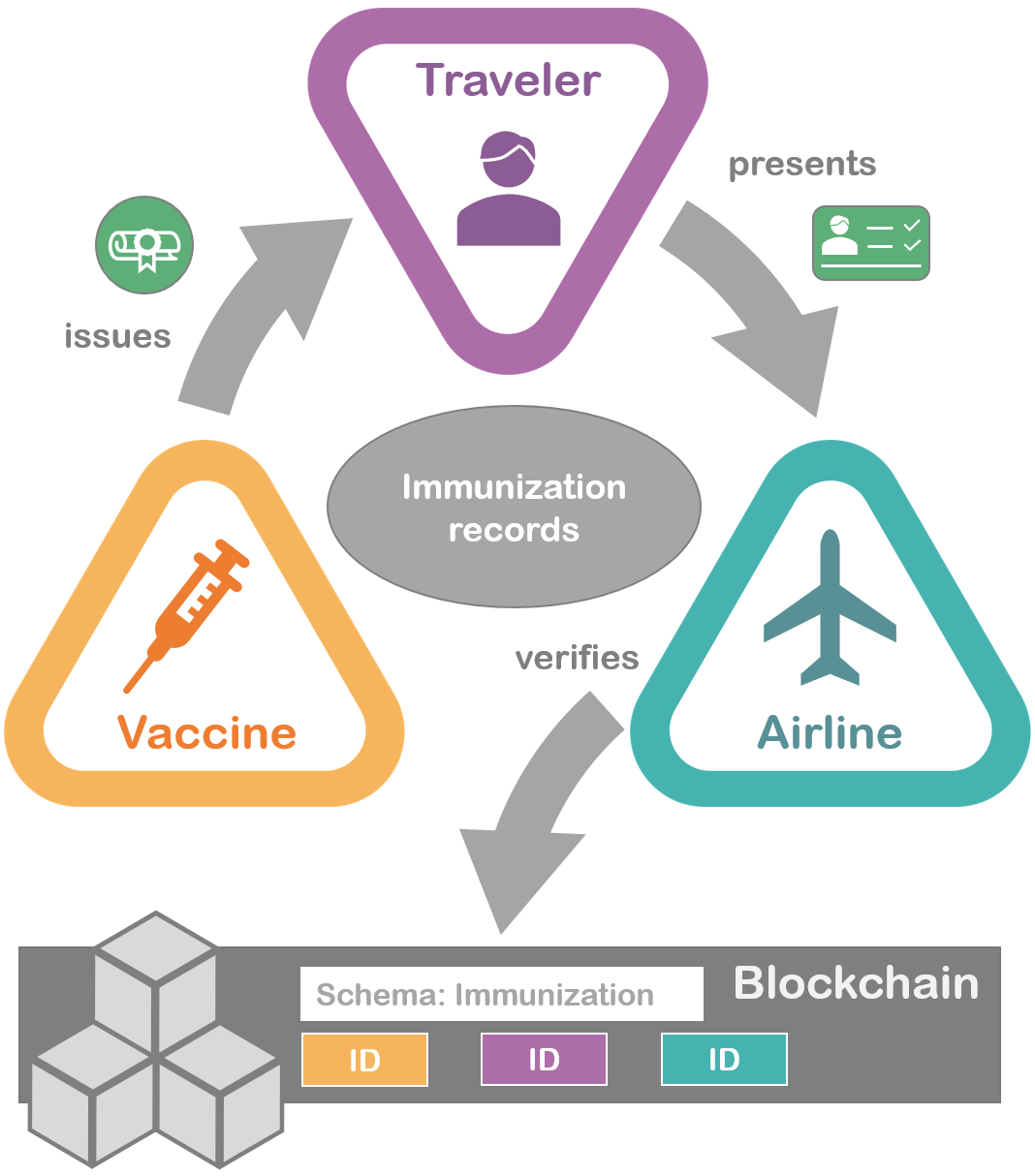}
  \end{center}
  \caption{Immunization records}
\end{wrapfigure}

\textbf{Context:} the COVID-19 pandemic highlighted the need for mass immunization to fight the spread of diseases. Even before the pandemic, proof of immunization was required for certain activities: traveling abroad, school enrollment, 
etc.

\textbf{Problem/Challenge:} despite previous attempts to create an international standard for proving immunization status, such as the International Certificate of Vaccination or Prophylaxis~\cite{icvp2005},
required when arriving from areas at risk of Yellow Fever, there is no easy way to validate the legitimacy of a person's reported immunization status. Reliable proof of vaccination is essential to support the 
regulations that, for instance, require students to be immunized to attend in-person classes at schools~\cite{alesp2020}.

\textbf{Storytelling:} Louise has a business trip. In order to board a flight to her destination, she must show proof of a SARS-CoV-2 immunization. She was vaccinated, and the proof of immunization has been recorded as a verifiable credential, by an accredited vaccination center, in her digital wallet. On the day of the travel, she goes to the airline counter and presents 
a QR code containing a verifiable record of her vaccination status. The airline employee scans the QR code and validates the immunization record, stating that Louise is ready to embark.

\textbf{Benefits:} the success of many efforts to bring society back to normal after the COVID-19 pandemic depends on reliable proof of vaccination, the so-called COVID passport. International Air Travel Association, for example, is adopting SSI technologies to provide a travel pass app~\cite{iata2021} that allows travelers to present verifiable proof of negative test results and vaccine shots taken without disclosing personal information. 

\subsection{Laboratory tests}

\begin{wrapfigure}{L}{0.3\textwidth}
 \begin{center}
    \includegraphics[width=0.30\textwidth]{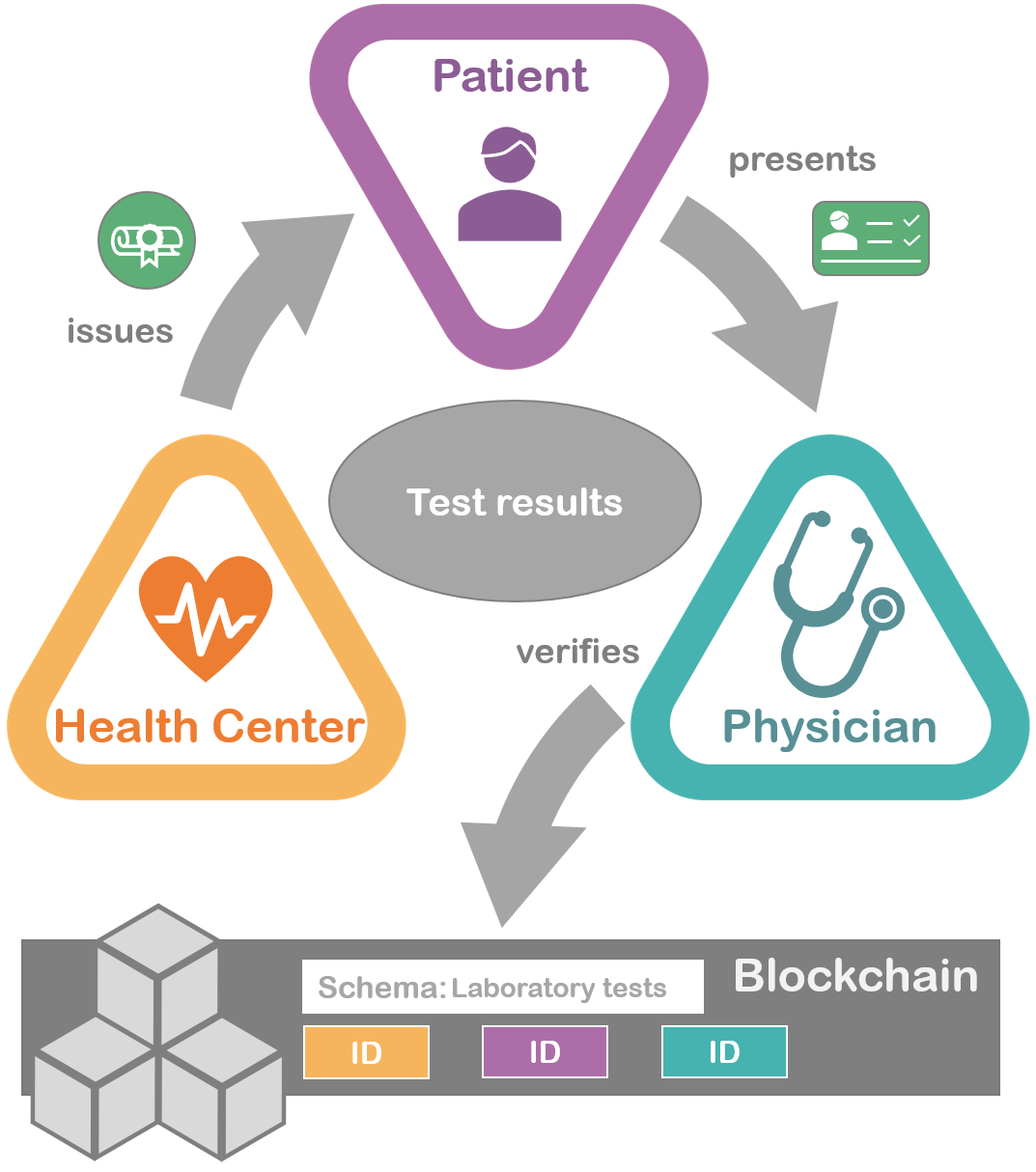}
  \end{center}
  \caption{Laboratory test results}
\end{wrapfigure}

\textbf{Context:} Lab tests examine human-derived samples to provide information for the diagnosis, prevention, treatment of disease, or health assessment of patients~\cite{uscdi2020}. Test results main contain reports, 
diagnostic images and videos, etc.

\textbf{Problem/Challenge:} the history of laboratory test results can provide valuable information for physicians regarding the evolution of patients' health conditions. However, labs usually keep the actual digital records of the tests and give the patients cumbersome printed copies of the results that tend to be discarded with time.

\textbf{Storytelling:} Mark is fighting liver cancer for many years. After two surgeries and many chemotherapy sessions, the disease appears to be under control. Every six months, however, Mark follows up with his oncologist to check his condition and have routine checkups. 

During the appointment, Mark's physician asks for the images from the last 3 MRI scans to compare with the most recent one. Mark recovers the lab results from his digital wallet and shares the files with the physician.

\textbf{Benefits:} patients should own their medical records: laboratories can issue test results to patients as verifiable credentials in an SSI digital wallet, allowing them to maintain the history of their private health information. Labs would benefit from reducing their liability as they would not have to keep confidential records in their possession.

\subsection{Control methods for prescription drugs}

\begin{wrapfigure}{L}{0.3\textwidth}
\begin{center}
    \includegraphics[width=0.30\textwidth]{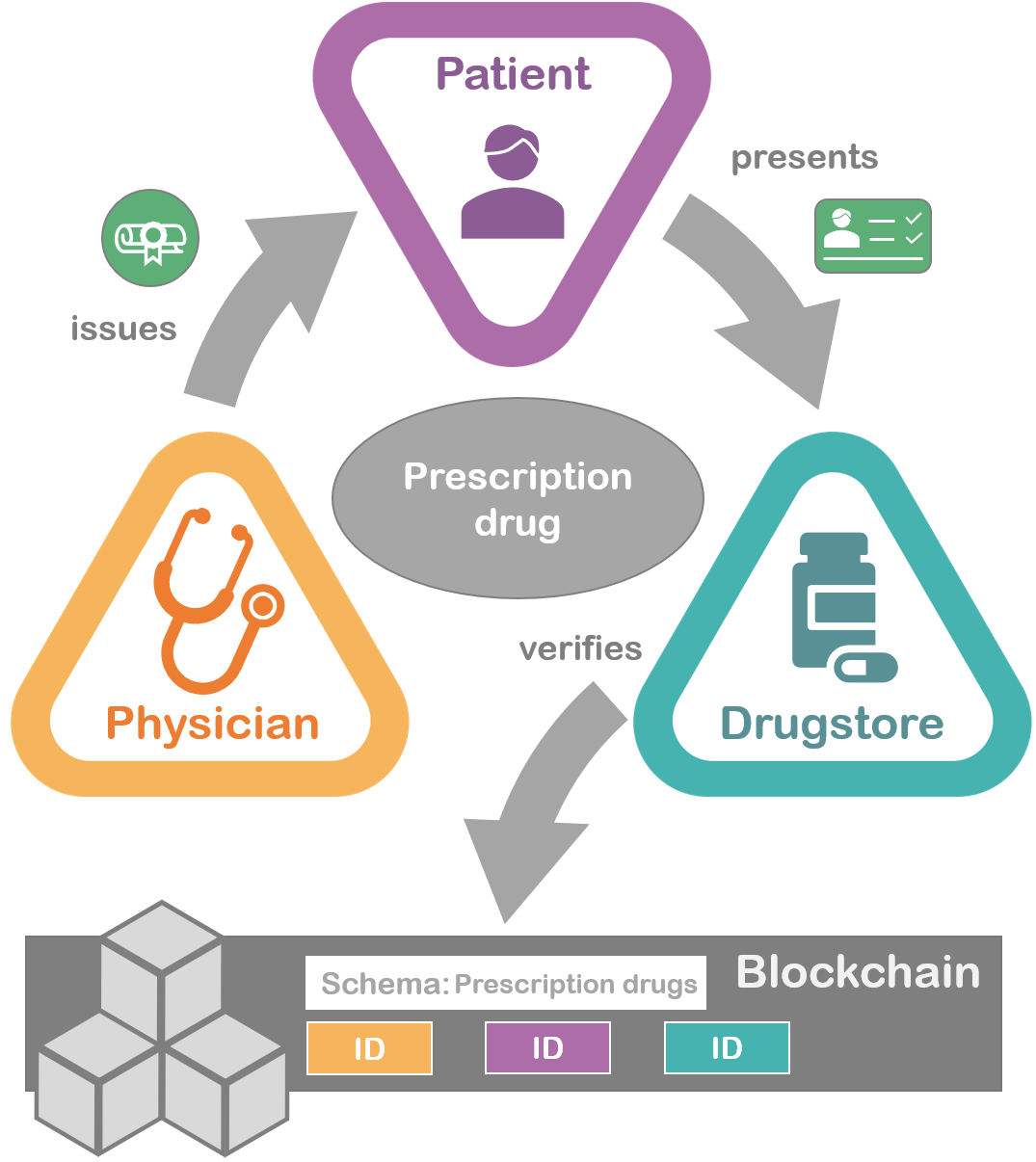}
  \end{center}
  \caption{Prescription drugs}
\end{wrapfigure}

\textbf{Context:} prescription drug abuse represents a health challenge worldwide. According to US~\cite{nida2016}, in 2014, more than 47.000 people died due to an unintentional drug overdose. 

\textbf{Problem/Challenge:} health agencies regulate control mechanisms for purchasing prescription medication, often requiring patients to present valid medical prescriptions. Among many difficulties to implement such controls, false prescriptions and identity forgery are the most common. Also, enforcing centralized systems to control medical prescriptions cause adoption friction in pharmacies and healthcare providers.

\textbf{Storytelling:} Brianna is having a hard time healing after a car accident. Her broken leg hurts. 
So, Brianna asks the doctor for pain relievers to help her. 
The doctor issues Brianna a prescription for a controlled painkiller, using his digital ID to sign the prescription straight to Brianna's digital wallet. 
To prove she can purchase the prescription drug, Brianna presents a QR Code to the drug store attendant containing a verifiable credential issued by her physician, with the drug designation, dosage, quantity, and posology. The attendant is able to confirm the presented information and proceed with the sale.

\textbf{Benefits:} prescription drug regulation can benefit from the SSI approach by leveraging blockchain features such as immutability and decentralized trust to combat counterfeit and identity fraud. In addition, selective disclosure allows patients to prove they are entitled to purchase prescription medicine without exposing their personal data.

\subsection{Medical procedures}

\begin{wrapfigure}{L}{0.3\textwidth} \begin{center}
    \includegraphics[width=0.30\textwidth]{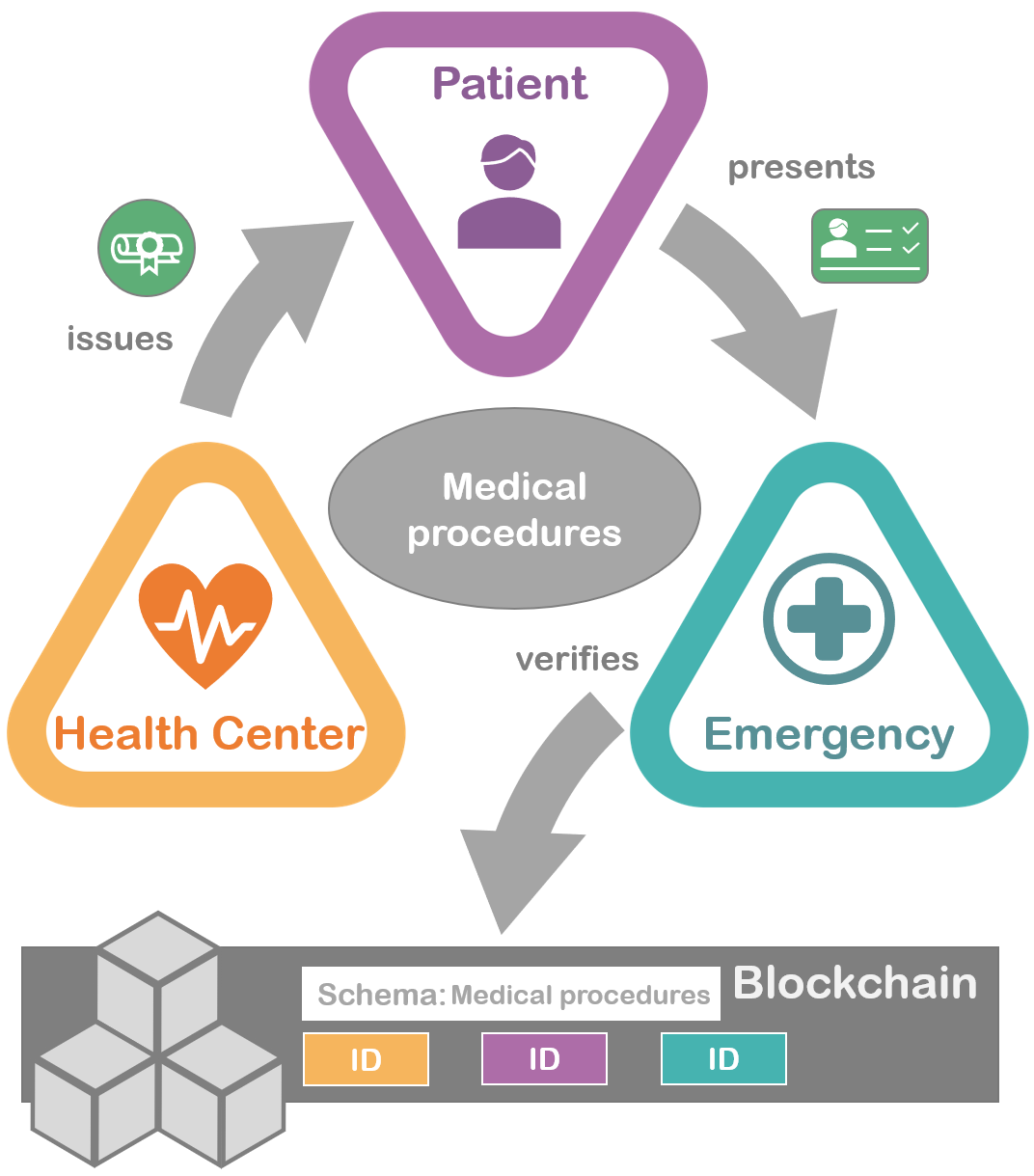}
  \end{center}
  \caption{Medical Procedures}
\end{wrapfigure}

\textbf{Context:} the ~\cite{uscdi2020} describes a procedure as ``an activity that is performed with or on a patient as part of the provision of care''. Examples of medical procedures include surgeries,  biopsies, amputations, general anesthesia, chemotherapy, and endoscopy.

\textbf{Problem/Challenge:} detailed reports on medical procedures are the type of information that is usually siloed, kept by the organizations that perform the procedures. Health professionals must be informed about  procedures undertaken by patients in order to offer proper treatment and avoid any health damage.

\textbf{Storytelling:} Sophia has had a stomachache for two days. She goes to the emergency room, tells the doctor about the symptoms, and points out where it hurts. She also recalls that she went through an endoscopy two days ago and started feeling bad since then. After assessing the endoscopic report Sophia shared electronically with him, the doctor speculates that she may be suffering from minor internal bleeding. The doctor then forwards her for further tests before prescribing the proper treatment.

\textbf{Benefits:} keeping medical records as credentials in an SSI digital wallet allow patients to break the information silos by creating a repository to aggregate reports on previous procedures, known medical conditions, etc.

\subsection{Vital signs}

\textbf{Context:} Vital signs are ``physiologic measurements of a patient that indicate the status of the body's life sustaining functions.''~\cite{uscdi2020}. This patient-generated health information can be collected from people in Intensive Care Units (ICUs) that require constant monitoring, but also from wearable devices that record health and lifestyle information. 

\textbf{Problem/Challenge:} from smartwatches to peacemakers, going through 
ICU medical monitors and glucometers, many devices can measure and collect vital signs. However, few devices follow any standard for structuring and storing this data, creating silos of health information that could produce valuable sources of information for physicians during medical evaluations.

\begin{wrapfigure}{L}{0.3\textwidth}
 \begin{center}
    \includegraphics[width=0.30\textwidth]{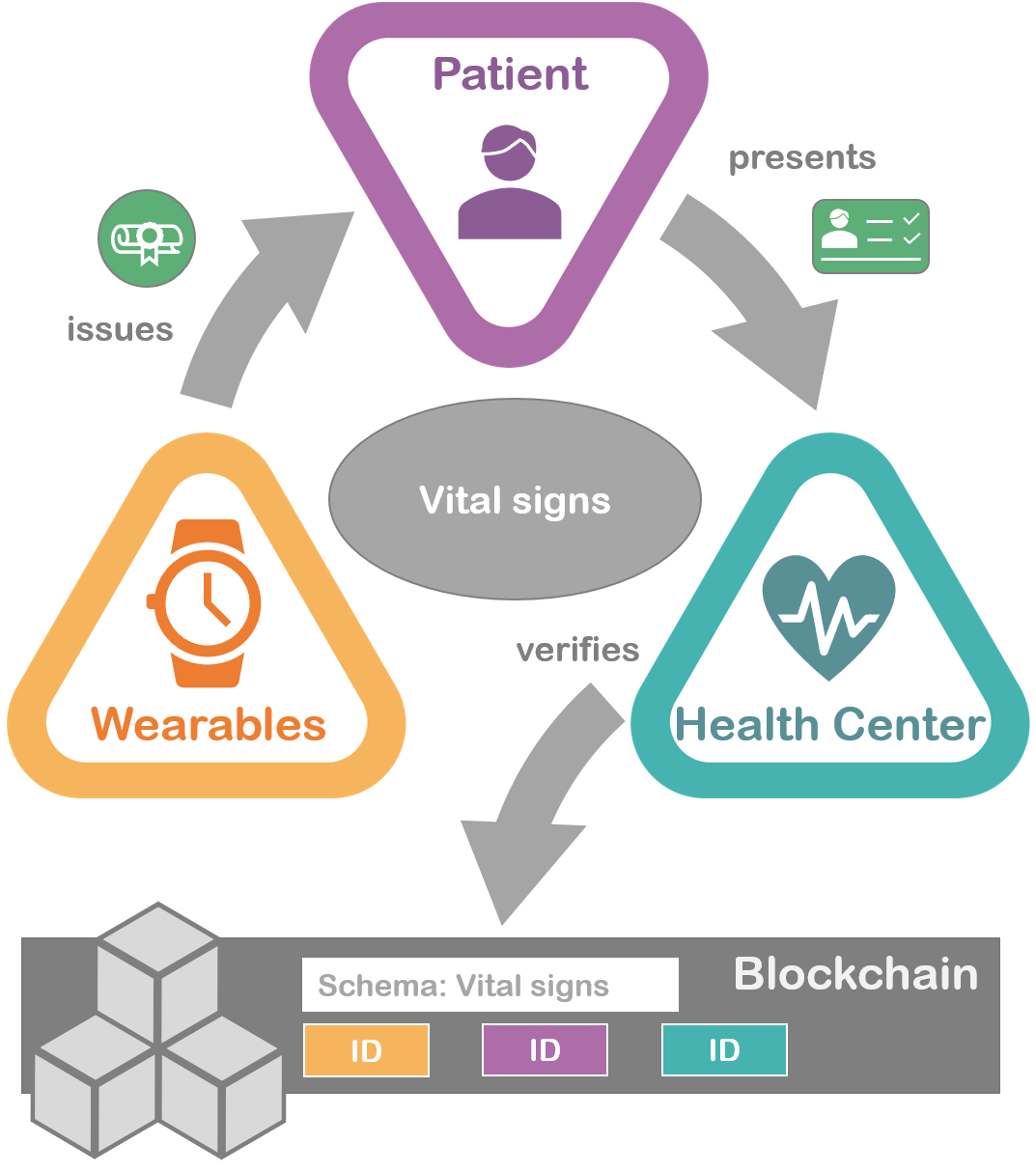}
  \end{center}
  \caption{Vital signs}
\end{wrapfigure}

\textbf{Storytelling:} Phil is late with his annual checkup. During the consultation, Phil's doctor orders routine screenings and a complete physical test. Phil recalls that his smartwatch has a heart rate and blood pressure monitor, and all these measurements are collected and stored 
in his digital wallet. Then, Phil uses it to generate a downloadable report of his vital signs from the past two months and share it with his doctor. This data, along with 
other reports, provides a rich data set that helps to assess his health condition. 

\textbf{Benefits:} vital signs express individual features the same way identity claims do. SSI credential schemas can be used to model biomedical data structures that support collecting vital signs using wearables, IoT medical devices, and regular medical monitors. UCIs immediately benefit from this technology, using SSI digital wallets as electronic health records for their inpatients.

\section{Architecture}
\label{arch}

\begin{figure}[t]
    \centering
    \includegraphics[width=0.74\textwidth]{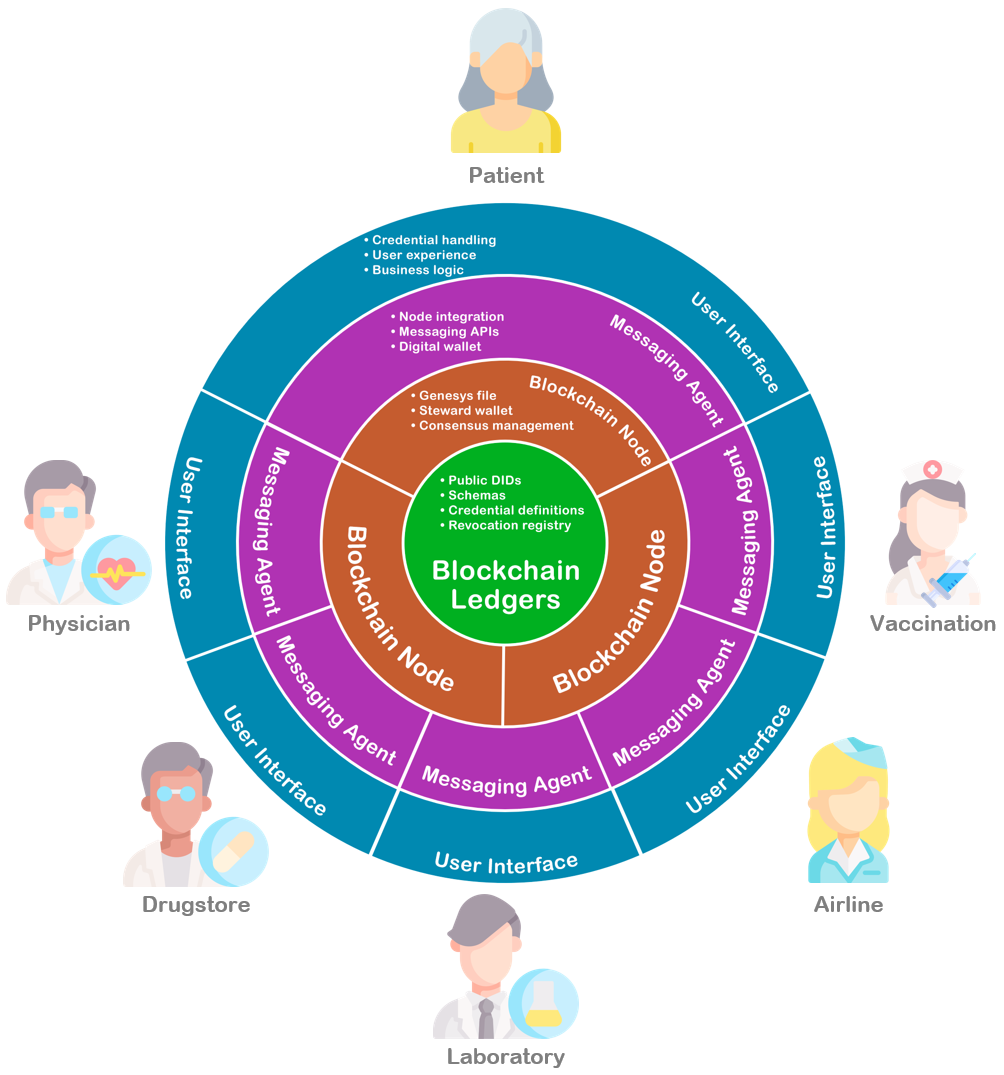}
    \caption{Context diagram representing layers and responsibilities}
    \label{fig:context}
\end{figure}

This section presents the proposed architecture of the blockchain-based self-sovereign health registry system. 
Figure~\ref{fig:context} portrays a context view, where the concentric circles depict the layers of the system.

The two inner layers are the foundation of the architecture. They represent the blockchain network and are common to all actors in the system. The innermost layer describes the \textbf{blockchain ledgers}, public information repositories that store data using a DLT. The blockchain ledgers hold the following information:

\begin{itemize}
    \item \textbf{Public DIDs:} identity records containing public keys and authentication parameters. Those DIDs enable verifiers to check if a presented claim is legit.
    \item \textbf{Schemas:} list of attributes used as data models for the credentials.
    \item \textbf{Credential definitions:} instances of schemas created for a particular issuer. It hold cryptographic hashes, bound to the issuer, for every attribute described in the schema. These definitions 
    allow that verifiers can check the validity of individual claims without disclosing the complete information.
    \item \textbf{Revocation registry:} even though ledgers do not store private credentials (holders keep those in their digital wallets), there must be a repository for revoked credentials where verifiers can assert if a presented credential is still valid.
\end{itemize}

The layer that surrounds the ledgers represents the \textbf{blockchain nodes}. Those nodes are the machines that support the blockchain network, managing the data that is stored in the ledgers. The \textbf{consensus management} algorithms ensure that every node has a synchronized copy of the ledgers. There can be as many nodes as needed to keep the blockchain network running efficiently, but the initial nodes are known as ``Genesis nodes''. Those initial nodes provide a unique identification of the network and are described in a file called \textbf{Genesis file}. The nodes and the messaging agents (explained below) must hold this file to connect to the network. The nodes are responsible for writing and reading data from the ledgers, and therefore, they need their own DIDs. 
Part of the node's DIDs is stored in their digital wallets, named \textbf{steward wallets}.

The \textbf{messaging agent} layer is responsible for holding \textbf{digital wallets} encrypted and protected, and for connecting with others blockchain nodes and agents. These two different connections are essential for understanding how information flows among entities and are explained in detail in the following sections. In summary, those connections can be described:

\begin{itemize}
    \item \textbf{Node communication:} using HTTP protocol, the agents connect to the blockchain network to write and read public data from the ledgers. The agents also expose \textbf{messaging APIs} to user interface components for credential handling.
    \item \textbf{DIDComm:} a protocol for secure peer-to-peer communication between agents, DIDComm allows entity agents to establish trusted relationships and exchange private messages and credentials. 
\end{itemize}


The outermost layer represents the \textbf{user interface}, where specific \textbf{business logic} is implemented. The user interface provides entity users with \textbf{credential handling} capabilities, such as credential issuance, credential verification, and relationship establishment. In straight terms, the user interface layer allows users to interact with the system. 

\subsection{Actors and interactions}

Once the overview of the architectural layers is presented, it is essential to describe the actors and their roles in the system. 
The diagrams in Figure~\ref{fig:uc:1} depict the use cases covered by the proposed architecture and the interactions among actors.

\subsubsection{Issuer/holder interactions}

\begin{figure}[t]
 \subfigure[Interaction Issuer/Holder]{
           \label{fig:usecase issuer/holder}
          \includegraphics[width=.49\textwidth]{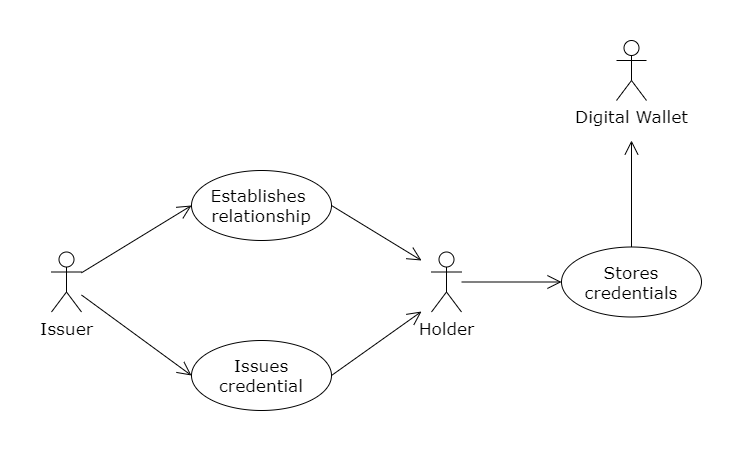}}
 \subfigure[Interaction Issuer/blockchain]{
          \label{fig:usecase issuer/blockchain}
          \includegraphics[width=.49\textwidth]{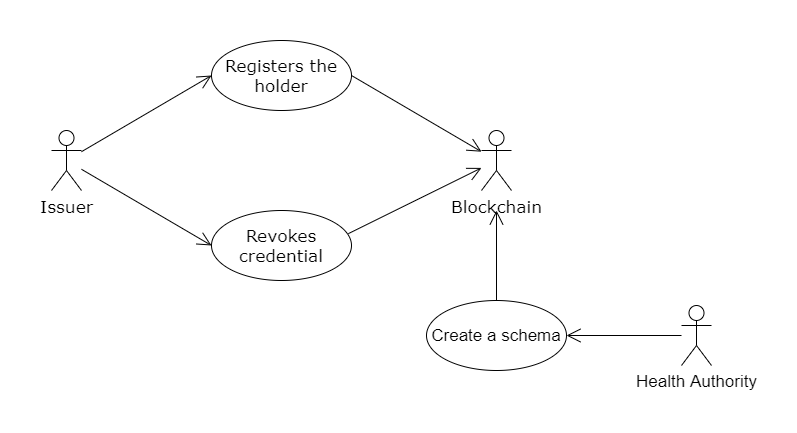}}
\caption{Use case diagram.}
\label{fig:uc:1}
\end{figure}

Issuers and holders are actors that represent the roles of entities in a SSI context. In the proposed architecture, their interaction is depicted in Figure~\ref{fig:usecase issuer/holder}.


The ``issues credential'' use case refers to a generic capability that describes the act of an issuer that asserts claims about a holder through the issuance of verifiable credentials. The system allows this use case to be extended to more specific scenarios, covering detailed interactions among specialized versions of the actors. An issuer also ``establishes relationship'' with holders before issuing credentials. This use case describes the exchange of DIDs between the issuer and holder. These DIDs and the issued credentials are stored in the holder's digital wallet (``store credentials'' use case).

As part of the issuer/holder interactions, there are use cases that illustrate actions that issuers perform in the blockchain. These use cases are depicted in Figure~\ref{fig:usecase issuer/blockchain}.


A peripheral use case ``create a schema'' is also represented in the diagram: to issue a credential, the issuer must reference an existing schema. Schemas must be registered in the blockchain by health authorities, restricting the number of available schemas according to the capabilities implemented in the system.

\begin{figure}[t]
    \centering
    \includegraphics[width=0.88\textwidth]{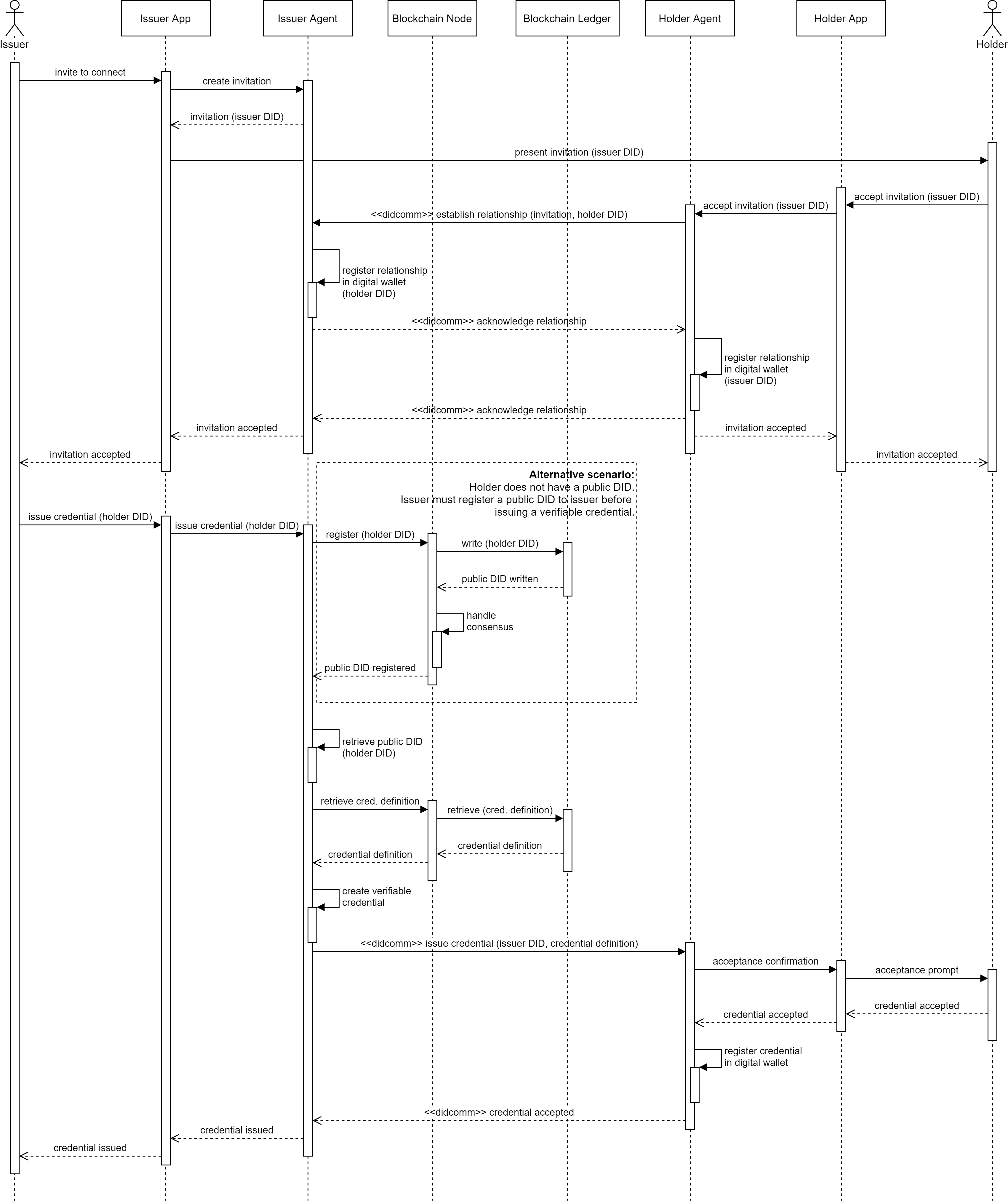}
    \caption{Sequence diagram - interactions issuer/holder}
    \label{fig:sequence credential issuance}
\end{figure}

Figure~\ref{fig:sequence credential issuance} depicts a detailed sequence diagram that explains the interaction between issuers and holders. In a step-by-step representation: an issuer (e.g., a physician) wants to issue a credential (e.g., medical prescription for a controlled drug) to a holder (e.g., a patient) and this involves establishing a relationship, exchanging digital IDs and transferring the credential use cases as mentioned earlier. 

\subsubsection{Verifier/holder interactions}

\begin{figure}[t]
 \subfigure[Interaction holder/verifier]{
          \label{fig:usecase holder/verifier}
          \includegraphics[width=.49\textwidth]{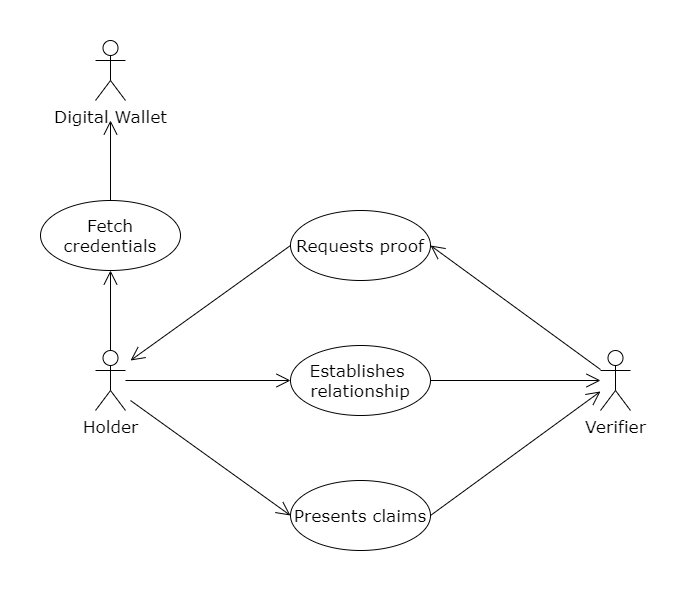}}
 \subfigure[Interaction verifier/blockchain]{
          \label{fig:usecase verifier/blockchain}
          \includegraphics[width=.49\textwidth]{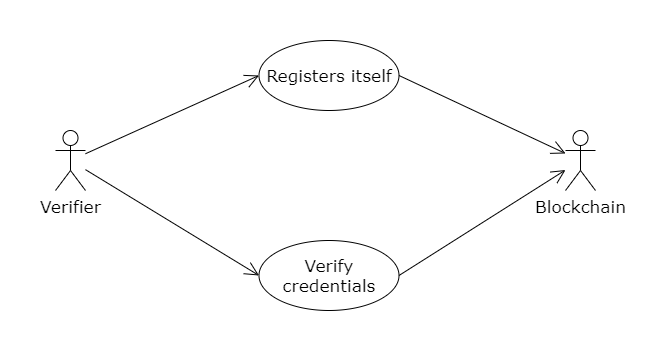}}
\caption{Use case diagram.}
\label{fig:uc:2}
\end{figure}

Verifiers and holders are actors that also represent the roles of entities in a SSI context. In the proposed architecture, their interaction is depicted in Figure~\ref{fig:usecase holder/verifier}.                                                                                                        


The ``present claims'' use case  refers to a generic capability that describes the act of an issuer that present its claims to a verifier. The system allows this use case to be extended to more specific scenarios, covering detailed interactions among specialized versions of the actors. A holder ``establishes relationship'' with verifiers before presenting their credentials. This use case describes the exchange of DIDs between the verifier and holder. Once the relationship is established, the verifier can ``request proof'' from issuers, selecting which attributes the holder must present. The holder then ``fetches the credentials'' that match the criteria defined by the verifier to create the credential presentation.

As part of the verifier/holder interactions, there are use cases that illustrate actions that verifier perform in the blockchain. These use cases are depicted in Figure~\ref{fig:usecase verifier/blockchain}.


Verifiers perform credential validation through ``verify credentials'' use case, fetching the holder an issuer public DIDs to evaluate the presented cryptographic proofs. 


\subsection{Components and adopted technologies}

\begin{figure}[t]
    \centering
    \includegraphics[width=0.9\textwidth]{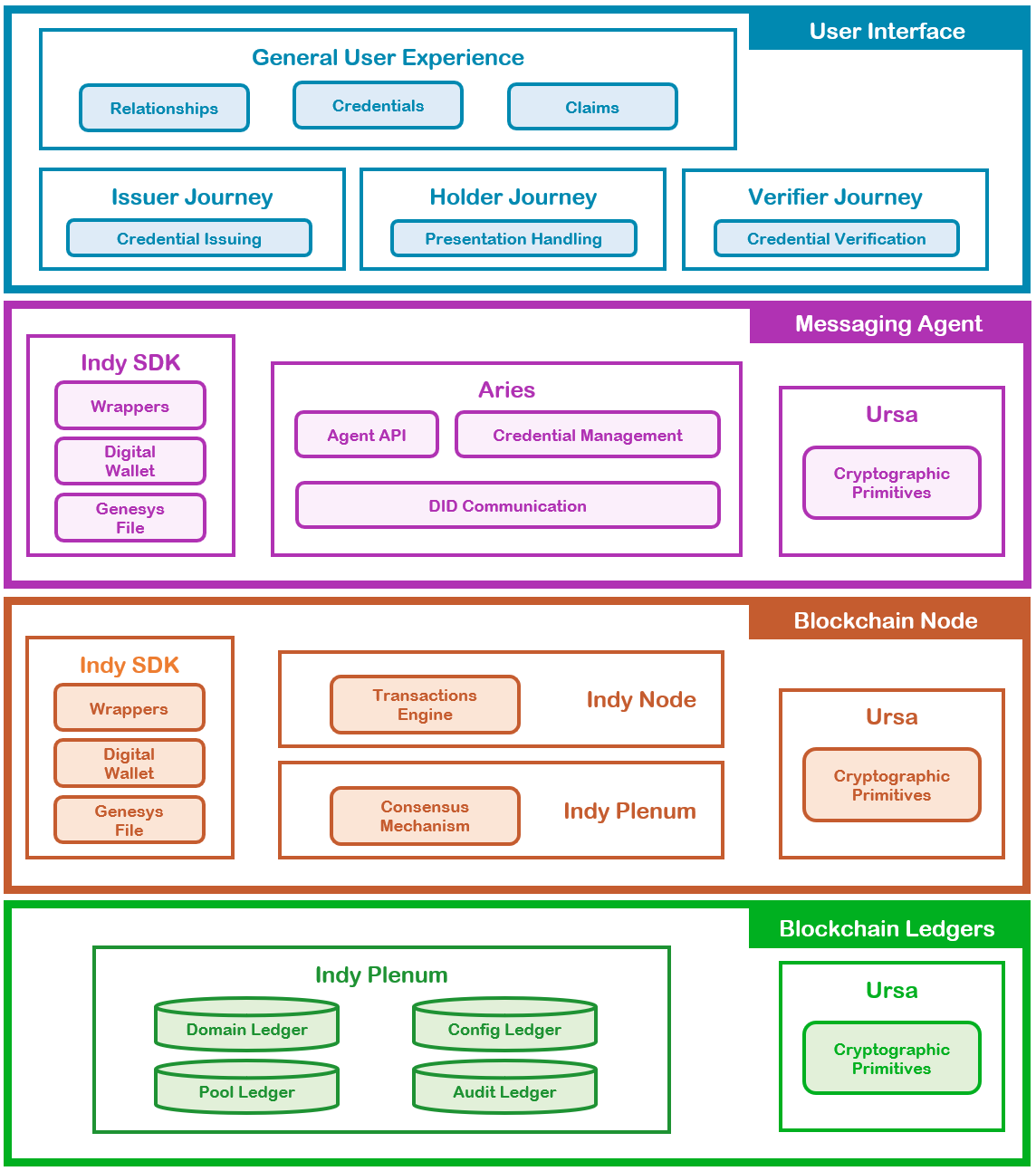}
    \caption{Architectural layers and stacked software packages}
    \label{fig:component diagram}
\end{figure}

The proposed architecture is comprised of several software components. Figure~\ref{fig:component diagram} depicts a diagram that groups these components in software packages and distributes them in the architectural layers shown in Figure~\ref{fig:context}, represented as stacked blocks. 

The \textbf{Blockchain ledgers layer} uses Hyperleger Indy's Plenum module to handle the storage of public identity information in distributed ledgers and Hyperledger Ursa's library to perform the necessary cryptographic primitives.
Hyperledger Indy and Hyperledger Aries also reference Ursa's crypto library to reuse complex cryptographic primitives across Hyperledger projects. The \textbf{Blockchain node layer} runs over Hyperledger Indy's Node, Plenum and SDK modules, managing the blockchain network and its consensus mechanisms. 

The \textbf{Messaging agent layer} represents the clients of the Blockchain network and runs over the Hyperledger Aries agent and embedded Indy SDK module. The \textbf{User interface layer} is an abstraction for the healthcare software that uses Messaging agent APIs to provide specific business journeys for Issuers, Holders, and Verifiers.

\section{Discussion}
\label{discuss}

We implemented the proposed architecture for validation, using the Indy/Aries architecture, installed on a cloud server, and the Trinsic wallet\footnote{\url{https://trinsic.id}}.

The main paradigm shift in system design is the ownership of data, which are under the control of the patient. This reduces the amount of data held by third parties, therefore reducing the potential for data leakage. However, it is possible to highlight new challenges.

The first challenge is \textbf{client-side storage}. In an ideal scenario, this storage should be in a local repository (e.g.,  patient's cell phone) with automatic backup to the cloud, similar to how Dropbox currently operates. 
The techniques to manage secure local writing and automatic cloud backup must be developed. In addition, we have to find out whether the storage medium on a modern cell phone is adequate for storing basic health data. Of course, some data, such as images and videos, should only be stored in the cloud.

Another challenge is the \textbf{usability} of the system. We believe that soon, the population will use services based on reading QRCodes and digital signatures. Still, the ease of using these interfaces should also be a determining factor for its wide adoption. An important line of research is to make the exchange of health data between Issuer and Holder automatic --- or almost automatic, especially when data collection takes place through sensors of Internet of Things.

A third challenge is the \textbf{maintenance} of the computing infrastructure. It is clear that patients must maintain local nodes, and Issuers and Verifiers must maintain messaging agents. But who should support the blockchain Indy nodes? The business model is not well defined.

Finally, we believe there is a significant economic potential in developing software, as each entity, Issuer or Verifier, may need to customize their messaging agents to meet their business. We also believe in the emergence of new business chains 
based on the trustworthiness of credentials and data availability. For example, a heart test could be used to enroll in a gym; attendance at the gym can generate discounts on health insurance; health coverage can create discounts on life insurance, and so on.

\section{Conclusions}
\label{conclusions}

This work exploits the main applications of SSI in healthcare scenarios. 
In our proposal, all data is owned by the patients. And it relies on SSI, blockchain technology and well-known standards. 

The architecture is flexible in order to satisfy several standardized uses cases. It does not solve dilemmas related to data accessibility versus privacy, but it provides appropriate means for the explicit consideration of each of these issues.

For the future, we plan to implement a functional prototype of the proposed architecture, shown in Section~\ref{arch}, and obtain feedback from patients and healthcare professionals and institutions about the usability of the system.
In addition, after to validate the basic solution, we could create more complex data health management models.

\bibliographystyle{sbc}
\bibliography{sbc-template}

\end{document}